\documentclass[aps,reprint,showkeys,noshowpacs]{revtex4-1}
\usepackage[utf8x]{inputenc}
\usepackage{amsmath,amssymb,url,graphicx}
\begin{document}
\title{Trend prediction in temporal bipartite networks:\\
the case of Movielens, Netflix, and Digg}
\author{An Zeng, Stanislao Gualdi, Mat\'u\v s Medo, Yi-Cheng Zhang}
\affiliation{Physics Department, University of Fribourg, CH-1700 Fribourg, Switzerland}
\date{\today}

\begin{abstract}
Online systems where users purchase or collect items of some kind can
be effectively represented by temporal bipartite networks where both
nodes and links are added with time. We use this representation to
predict which items might become popular in the near future. Various
prediction methods are evaluated on three distinct datasets originating
from popular online services (Movielens, Netflix, and Digg). We show
that the prediction performance can be further enhanced if the user social
network is known and centrality of individual users in this network is
used to weight their actions.
\end{abstract}


\maketitle

\section{Introduction}
Many websites give their users the possibility to buy, review or
simply share various kinds of products or other contents. This is
the case, for example, for e-commerce sites as Amazon or social
networks like Twitter and Digg. Data produced by systems of this kind
can be effectively described by bipartite networks which consist of
two types of nodes (representing users and items), and where every
edge runs between a user node and an item node if the user has
\emph{collected} (bought, rated, or otherwise favored) this item.
This representation has been successfully used to, for example,
assign reputation values to nodes in a network~\cite{Deng2009},
study global structural properties of interlocking company
directors~\cite{Robins2004}, and to compute personalized
recommendations by a random walk process on the network~\cite{Zhou07}.

In addition to collecting items, users can often make explicit links
to other users by, for example, following them (in which case items
collected by followed users are automatically forwarded to the users
who follow them). This gives rise to a monopartite network where
nodes are connected through directed edges representing leader/follower
relationships (as is the case for Twitter) or undirected edges
representing friendship relationships (as is the case for Facebook).
Thanks to the availability of large-scale data, online social networks
have been studied extensively. Their network characteristics have been
measured~\cite{Mislove2007} and compared with those of real-life social
networks~\cite{Ahn2007}, statistical analysis led to various models of
their evolution~\cite{Leskovec2008,Kumar2010}, and studies of user
influence aimed at finding influential spreaders~\cite{Kempe2003,Lu2011}
or extracting the subset of user-user connections that actually drive
user behavior~\cite{Huberman2008}.

The bipartite and monopartite network are often connected not only by
sharing the same set of users but they also exert mutual influence:
a link between two users influences items collected
by them in the future and collecting similar items may lead to a pair
of users being aware of each other and eventually connected by a link.
While much of the work so far assumed a static picture where a particular
snapshot of a network is studied without considering time when individual
edges were created, increased availability
of datasets with time information allows for a better perspective on
network formation and function (see~\cite{Holme2011} for a review). Time
labels are of particular importance in information networks~\cite{Medo2011}
where they can be used to design specialized information filtering algorithms
that can account for changing interests of users~\cite{Ding2006} or prefer
recent scientific publications over old ones~\cite{CiteRank07}.

Once time labels become available, it is natural to ask how well are
we able to predict the future development of a given network. This is a
very practical question as the ability of making good predictions of
future popularity of items is important for vendors and their
marketing strategies. Knowing potential hot items is of great interest
also to users who want to avoid plowing through the bulk of mediocre items.
Site administrators can benefit from this kind of knowledge too
because it allows them to better use their system resources. For
example, a video which is flagged to potentially become very popular in the near future
can be made available for download from multiple mirrors of the web site.
Predicting future trends is relevant also from a theoretical point of
view since the individuation of informative signals may help to
isolate the basic mechanisms driving the network's evolution and eventually
contribute to the understanding of complex connection patterns in
real networks. There have been works that studied the time evolution
of popularity of online content without considering the question of
prediction. An extensive study of various temporal patterns of popularity
in online media has been presented in~\cite{Yang2011}. Similarly, a
classification of YouTube videos into three classes according to how
their popularity decays after an initial burst was reported
in~\cite{Crane2008} and supplemented by a model of user behavior
combining an epidemics-like propagation of interesting content and
a power-law distribution of waiting times.

Predicting the user interest prior to publication of items based only
on item features turns out to be very
difficult~\cite{Tsagkias2009}. The situation becomes much
different after publication when robust patterns develop
fast. In the case of the popular online service Digg.com where users
submit links to stories and comment on stories submitted by the others,
predictive models trained on comments
made in the first few hours after the story submission can
successfully predict its later popularity~\cite{Jamali2009}. Much
more information is hidden in the initial growth popularity: the
popularity of a story as early as one hour after its submission has
been shown to correlate strongly with its final
popularity~\cite{Szabo2010}. On the other hand, the same authors
report a lower level of predictability for YouTube videos which they
attribute to much longer time scales and a lack of popularity
saturation there. An explicit popularity evolution
model based on how Digg users can reach the site's content was
presented in~\cite{Lerman2010}. There are several other suggestions
that predictions based on our actions in online environments can be
particularly effective thanks to their high level of automation
and exceptional coverage (online data can be collected and evaluated
automatically for millions of users from all parts
of the world). For example, Google aggregates search queries to track
the level of flu activity (see~\url{www.google.org/flutrends/}).
Another example is a study of Twitter mood as a predictor of the
stock market moves~\cite{Bollen2010}.

In this paper, we study three distinct datasets created by popular
online services---Netflix, Movielens, and Digg---with focus on
temporal patterns of user behavior and popularity prediction.
Different from~\cite{Szabo2010}, we do not follow individual items
after their submission. We focus instead on a given time point and
attempt to predict which items may become the most popular in a given
future time window. Rather than focusing directly on a particular
algorithm, we progress in steps from basic empirical observations
made on the chosen datasets to methods which possess some
predictive power. The paper consists of two parts. We first consider
the bipartite networks of Movielens, Netflix, and Digg and use this
information to predict item popularity. We then discuss the
case of Digg where in addition to the user-item data, we also have
the social user-user network which allows us to improve predictions
of an item's popularity by considering the social status of users
who have collected this item.

\section{Trend Prediction in Bipartite User-Item Networks}
As testing data, we use datasets produced by three
popular online services: Netflix, Movielens, and Digg. The Digg
dataset has been obtained by the authors of~\cite{Lerman2010b} who studied
spreading of stories in social news sites. The dataset contains
information about stories promoted to the Digg's front
page in June 2009. For each story, it collects the list of all users
who have ``dug'' the story (voted for it) up to the time of data
collection (5th July 2009) and the time stamp of each vote. We also
retrieved the voters' friendship links within Digg.com. Our Netflix
data is based on the dataset released by the company for the
Netflixprize (see \url{www.netflixprize.com}). The original data has
$480,189$ users, $17,770$ items and $100,480,507$ ratings. Finally,
the Movielens data is based on the dataset with $10,000,054$ ratings
from $71,567$ users for $10,681$ movies. It has been released by the
GroupLens research group (see \url{www.grouplens.org/node/73}). Since
the original Movielens and Netflix datasets are large, we construct a
subset for each of them by randomly choosing $U$ users who have rated
at least $20$ movies and keeping all the movies that they rated.

In our user-item bipartite networks, we label users by Latin letters
and items (movies in Movielens or Netflix and stories in Digg) by
Greek letters. All datasets are mapped into an adjacency matrix
$\mathsf{A}$ whose elements $A_{i\alpha}$ are equal to $1$ if user $i$
has collected item $\alpha$ and $0$ otherwise. In Digg, an item is
collected by a user if this user ``dug'' (gave their vote) the
item. In Movielens and Netflix, we have more complete information
consisting of review rating in an integer or half-integer scale from 1
to 5 which is then mapped to our binary data by applying a threshold
rating of $3$: any item rated $3$ or above is marked as
collected by a respective user. The number of users $U$, items $I$ and
resulting links $L$ together with the time period when the data was
collected are given in Table~\ref{tab_dataset}. Compared
to the original data, the subset contains about $7\%$ of users, $70\%$
of movies and $9\%$ of links for Movielens and $1\%$ of users, $90\%$
of movies and $12\%$ of links for Netflix. Since only temporal patterns
of individual items contribute to our popularity predictions in these
two datasets, one can expect that thus-created subsets have no effect
on these predictions.

\begin{table*}
\centering
\begin{ruledtabular}
\begin{tabular}{lrrrrr}
  Dataset &     $U$ &    $I$ &             $L$ & Start date             & End date\\
\cline{1-6}
Movielens &   5,000 &  7,533 & $8.6\cdot 10^5$ &  $1^{\rm st}$ Jan 2002 & $1^{\rm st}$ Jan 2005\\
  Netflix &   4,968 & 16,331 & $1.2\cdot 10^7$ &  $1^{\rm st}$ Jan 2000 & $31^{\rm st}$ Dec 2005\\
     Digg & 336,225 &  3,553 & $3.0\cdot 10^6$ & $31^{\rm st}$ May 2009 & $5^{\rm th}$ Jul 2009\\
\end{tabular}
\end{ruledtabular}
\caption{Basic properties of the used datasets: The number of users $U$,
items $I$, and links $L$ and the start/end date of the data collection.}
\label{tab_dataset}
\end{table*}

We consider snapshots of these networks at different time steps by
considering only the links established before given time $t$.
The time-dependent adjacency matrix $\mathsf{A}(t)$ then can
be used to introduce user degree
$k_i(t)=\sum_\alpha A_{i\alpha}(t)$ and item degree
$k_{\alpha}(t)=\sum_i A_{i\alpha}(t)$ which correspond to the number
of items collected by user $i$ and the number of users who collected
item $\alpha$, respectively. The popularity increase of item
$\alpha$ in past $T_P$ time steps (the past time window) is then
\begin{equation}
\label{increase}
\Delta k_{\alpha}(t,T_P) =  k_{\alpha}(t) - k_{\alpha}(t-T_P).
\end{equation}
For a suitably chosen value of $T_P$, this quantity measures
recent interest in item $\alpha$ (while a too small value leads to a
high noise level and many items with zero degree increase, a too high
value puts large weight on outdated developments at the expense of
recent changes). Note that when we speak about popularity in this
paper, we mean the absolute/total popularity (\emph{i.e.}, the current degree
$k$ of an item). When, instead, we speak about recent popularity or
popularity increase, we mean the degree increase $\Delta k$ of an item as in
Eq.~(\ref{increase}).

Our main goal here is to predict which items are expected to attract
the biggest attention in the near future. To this end we define a test
date $t^*$ and a future time window of length $T_F$, and rank all
items according to their popularity increase $\Delta
k_{\alpha}(t^*+T_F,T_F)$. We refer to this ranking as the \emph{true
ranking}. We then consider a generic predictor which, based on links
existing before time $t^*$, assigns scores $\{s_\alpha\}$ to all items.
These scores are then mapped into a \emph{predicted ranking}. To test
the performance of a predictor, we compute the fraction of items in
the top $n$ places of the estimated ranking that appear also in the
top $n$ places of the true ranking. This standard information
retrieval metric is called precision~\cite{Herlocker2004} and lies in
the range $[0,1]$ (the higher the better). We label it as
$P_n$ here. To obtain the final
evaluation of the performance of the predictor, we average results
over $9$, $12$, and $7$ regularly-spaced test dates $t^*$ for
Movielens, Netflix, and Digg, respectively. It is often the case
that items popular in the future time window $(t^*,t^*+T_F]$ were already
popular in the past time window $(t^*-T_P,t^*]$. Successful prediction
of those items, albeit contributing to precision $P_n$, brings smaller
benefit to the users than prediction of genuinely ``new entries'': items
that were missing in top $n$ in the past time window but they appear
there in the future time window. We label the true number of those items
as $E_n$ and the number of those successfully identified by our ranking
as $C_n$. The rate of predicting these new entries, $Q_n:=C_n/E_n$, then
allows us to measure how well a method is able to anticipate future trends
which are not yet obvious. While we always present results for top $n=100$
items, we evaluated prediction performance of all studied methods
also for other values ($50$ and $200$) and found that despite
the absolute values of $P_n$ and $Q_n$ change, the relative
comparison of the methods and our main conclusions still apply.

\subsection{Popularity-based predictors}
Preferential attachment, also known as the rich-get-richer process,
cumulative advantage, the Yule process, or the Matthew effect, is a
well-known mechanism of network evolution which assumes that the rate
at which nodes attract new links is proportional to their degree. In
our context this means that items that are popular at time $t^*$ are
expected to have better chances to attract new users, implying that
the current degree of an item $k_{\alpha}(t^*)$ is a good predictor
of its future popularity increase. Preferential attachment-based models
have been successfully used to explain the emergence of scale-free
structures in different systems, ranging from the World Wide
Web~\cite{BarAlb99} to
the number of species in a genus~\cite{Yule25} and scientific
citations~\cite{Price1976}. Despite its success, pure
preferential attachment is often too crude to reproduce a more
detailed behavior of real networks. In particular, it is often the
case that the interest towards individual items vanishes with
time~\cite{Medo2011} and the current degree thus becomes a poor
indicator of the future popularity increase. This is especially
true for information networks---a class to which belong all
three datasets studied in this paper.

\begin{figure*}
\centering
\includegraphics[width=2\columnwidth]{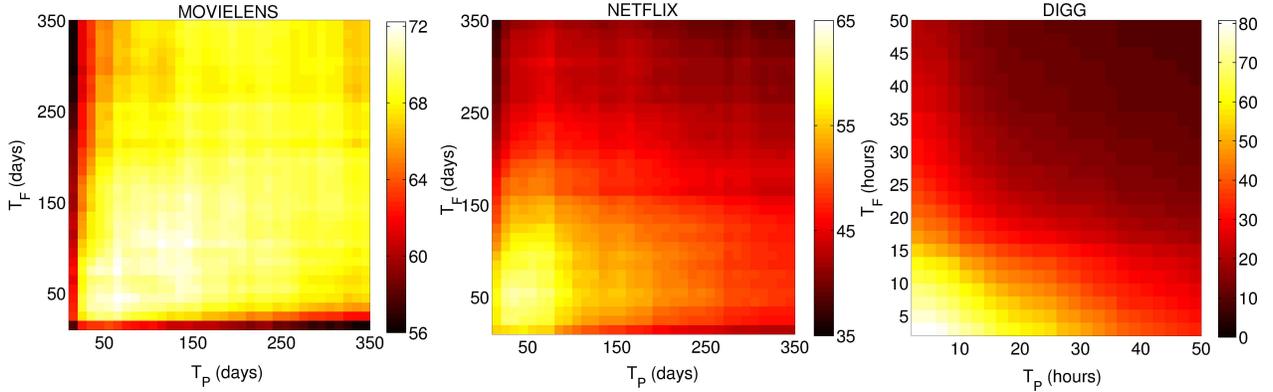}
\caption{(Color online) Heat map of precision $P_{100}$ obtained with
the degree increase predictor in the past/future time window plane
$(T_P,T_F)$. Time is measured in days for Netflix and Movielens and
in hours for Digg.}
\label{fig1}
\end{figure*}

To avoid the problem of decaying interest, one can base the
prediction on the probability of acquiring new links measured by the
recent popularity of an item. Assuming that in the future time
window this link-attracting probability does not change
significantly, the prediction score of an item at time $t^*$ can be
set as $\Delta k_\alpha(t^*,T_P)$ where $T_P$ is the length of the
time lapse in which the increase takes place. Fig.~\ref{fig1} shows
the prediction precision in the $(T_P,T_F)$ plane and demonstrates some
significant differences between the datasets. While both Movielens and
Netflix display optimal precision inside the plane, the popularity
increase decays very fast in Digg and, as a result, precision
decreases monotonically with both $T_P$ and $T_F$.
Since the $\Delta k_\alpha(t^*,T_P)$ predictor is simple and effective,
we use it as a benchmark for all later methods. As shown in Fig.~\ref{fig1},
$\Delta k_\alpha(t^*,T_P)$ generally performs best when $T_P=60\,\text{days}$
in Movielens, $T_P=60\,\text{days}$ in Netflix and $T_P=10\,\text{hours}$ in
Digg data. In the following analysis we always set $T_P$ to these values.

In the context of the Barab\'asi-Albert model, the expected
popularity increase of an item is proportional to the item's degree
and the two
predictors, $\Delta k$ and $k$, are expected to produce rankings that
are identical on average (though, $\Delta k$ is a more noisy
indicator than $k$). As already mentioned, patterns in real data often
substantially differ from the basic Barab\'asi-Albert
scenario and rankings produced by the two predictors are thus expected
to diverge to some extent. To benefit from these two complementary
sources of information, we introduce parameter $\lambda\in[0,1]$ which
interpolates between them and introduce the hybrid item score in the form
\begin{equation}
s_\alpha(t^*,T_P) = k_\alpha(t^*)-\lambda k_\alpha(t^*-T_P).
\label{pop_lam}
\end{equation}
This simplifies to the total popularity (degree) for $\lambda=0$ and
to the recent popularity (degree increase) for $\lambda=1$. We refer
to this as the popularity-based predictor (PBP).

\begin{figure*}
\centering
\includegraphics[scale=0.57]{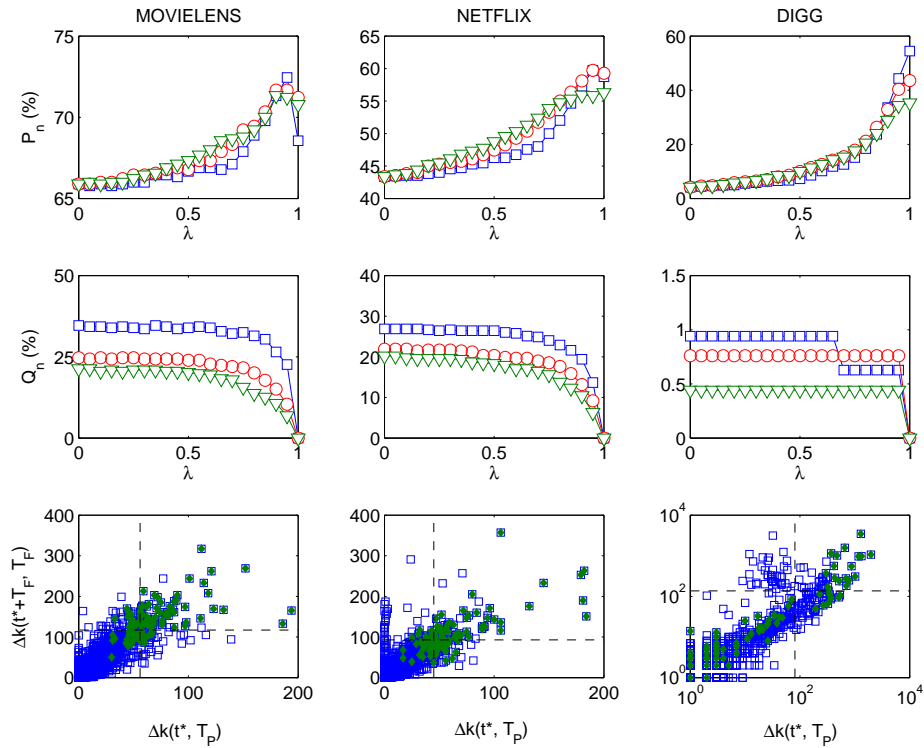}
\caption{(Color online) Results for the PBP predictor defined by
Eq.~(\ref{pop_lam}). $T_F$ is set to $90\,\mathrm{days}$ days for
Movielens/Netflix and $10\,\mathrm{hours}$ for Digg in all simulations.
\emph{Top row:} Precision versus $\lambda$ for different past time windows
$T_P$: blue lines with squares ($T_P=30\,\mathrm{days}$ in Netflix/Movielens
and $T_P=10\,\mathrm{hours}$ in Digg), red lines with circles
($T_P=60\,\mathrm{days}$ in Netflix/Movielens and
$T_P=20\,\mathrm{hours}$ in Digg), and green lines with triangles
($T_P=90\,\mathrm{days}$ in Netflix/Movielens and
$T_P=30\,\mathrm{hours}$ in Digg).
\emph{Middle row:} Rate of correctly guessed new entries
$Q_n$ versus $\lambda$ (different lines as in the top row).
\emph{Bottom row:} Scatter plots of item degree increase in
$(t^*-T_P,t^*]$ versus the degree increase in $(t^*,t^*+T_F]$ with
both history and future window lengths $60$ days for Netflix and Movielens
and $10$ hours for Digg.
Items placed by the PBP method with $\lambda=0.9$ in
top 100 are marked with full green symbols.}
\label{fig2}
\end{figure*}

Results obtained with PBP for different values of $\lambda$ are shown
in Fig.~\ref{fig2}. Recent popularity gives better results than total
popularity in all datasets, especially in Digg where interest in a
story fades quickly and the absolute popularity hence yields a
particularly low precision value. For both Movielens and Netflix,
there is an intermediate value of the parameter $\lambda$
outperforming both total and recent popularity. The optimal value of
$\lambda$ is approximately $0.9$ in both cases. The absence of such
a maximum in Digg confirms the intuition that the temporal evolution of
news popularity substantially differs from that of movies. The rate
of correctly predicted new items $Q_n$ monotonically decreases with
$\lambda$ in both Movielens and Netflix and reaches $0$ for
$\lambda=1$ (by definition because $\lambda=1$ corresponds to
prediction by popularity increase where items with low recent
popularity cannot score high). $Q_n$ values are very low in the case
of Digg which is due to the quick dynamics of news which makes it
nearly impossible that an old item with high degree can be among the
top growing items in the near future. This is in line
with~\cite{Szabo2010} where high correlation has been found between
popularity of stories early after their submission and their final
popularity.

Fig.~\ref{fig2} further includes scatter plots showing
popularity increase in the past and future time window for individual
items (no averaging over $t^*$ was applied).
The vertical dashed line marks the degree of the 100th
most popular item in the history window and the horizontal dashed line marks the
degree of the 100th most popular item in the future window. The
top $100$ items predicted by PBP with $\lambda=0.9$ are marked with
full symbols. The meaning of $C_n$ is
well illustrated by these scatter plots. Among the top $100$ most
popular items in the future time window, some were among the top
$100$ most popular also in the past time window (the
top-right quadrant in the scatter plots) and some are new---they
were not among the top $100$ most popular in the past (the top-left
quadrant). Items from the top-left quadrant are more difficult to be
predicted and for this reason they are more valuable. By setting
$\lambda<1$ in the PBP,
the top $100$ predicted items cease to be located only in the top-right
and bottom-right quadrant and some of them appear in our target top-left
quadrant ($C_n$ is equal to the number of these items) as well as in
the bottom-left quadrant (where they represent wrong predictions similarly
as the predicted items located in the bottom-right quadrant). We can
conclude that intermediate values of $\lambda$ highlight some of the
items that are increasing their popularity at a faster pace than they
did in the recent past. This is however not the case for Digg where
the value of $Q_n$ stays virtually zero regardless of $\lambda$.

\begin{figure*}
\centering
\includegraphics[scale=0.57]{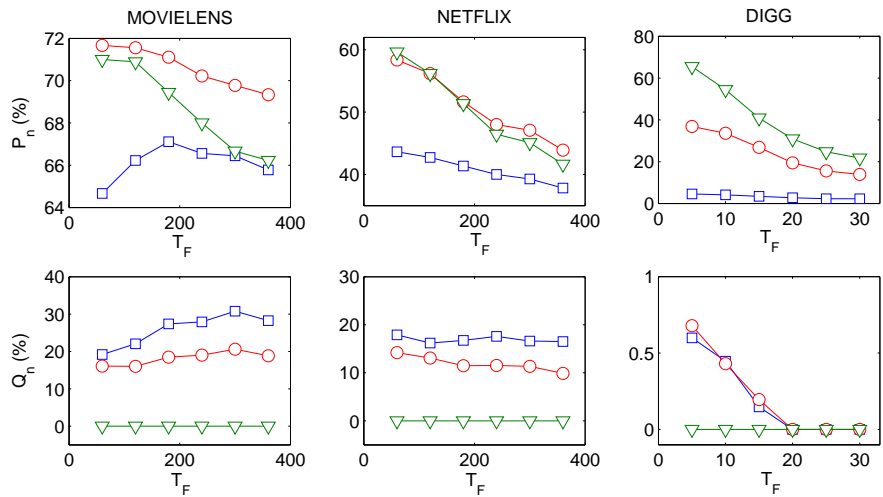}
\caption{(Color online) \emph{Top row:} Precision vs the future time
window length $T_F$ for different values of $\lambda$:
$\lambda=0$ (blue lines with squares), $\lambda=0.9$ (red lines with
circles), and $\lambda=1$ (green lines with triangles). The past time
window length is $T_P=2\,\mathrm{months}$ for Movielens/Netflix and
$10\,\mathrm{hours}$ for Digg.
\emph{Bottom row:} Rate of correctly guessed new entries $Q_n$
as a function of the future time window length $T_F$ and different
values of $\lambda$ (the same lines as in the top row).}
\label{fig3}
\end{figure*}

We further investigate how the PBP method performs as a function of the
future time window length $T_F$ for different values of $\lambda$. As
can be seen in Fig.~\ref{fig3}, predictions based on the popularity
increase ($\lambda=1$ and $\lambda=0.9$) give better results than
those based on total popularity ($\lambda=0$) but their
performance decreases with $T_F$ faster than for $\lambda=1$ and $\lambda=0.9$.
This confirms that total popularity is a reliable and stable predictor
for the long run but it can be outperformed by other methods
for short time windows. PBP with $\lambda=0.9$ gives on overall the best
performance and is rather stable when the future time window is varied.
In the case of Digg, it is always best to use pure popularity increase
for prediction and, closely related, the decrease of precision of with
$T_F$ is the steepest out of the three tested datasets.

\subsection{Trend setters: a weighted popularity predictor}
\begin{figure*}
\centering
\includegraphics[scale=0.57]{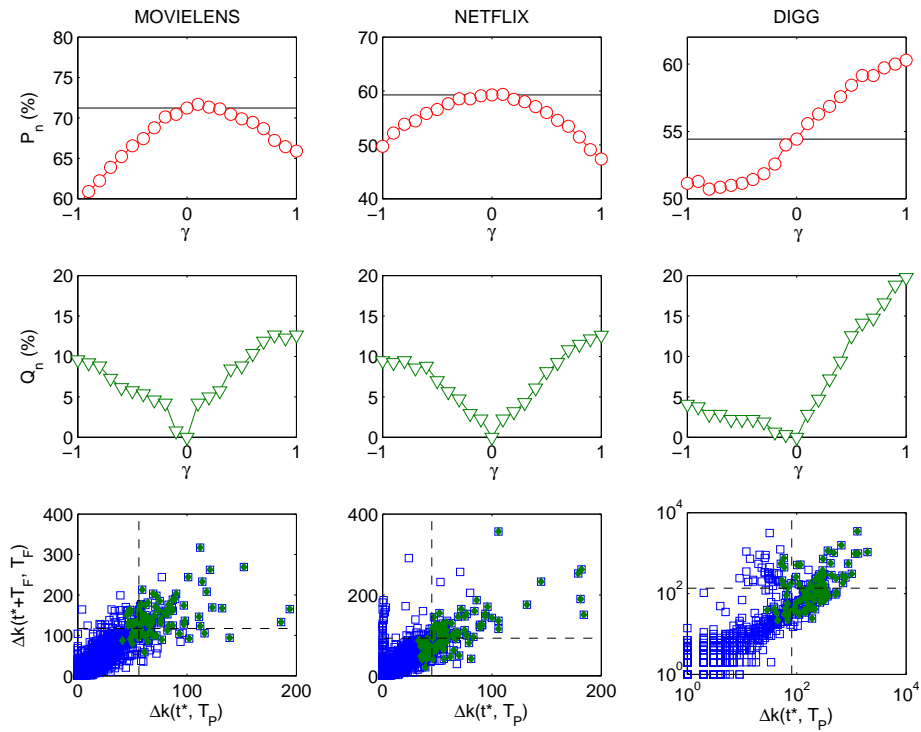}
\caption{(Color online) \emph{Top row:} Prediction precision versus
$\gamma$ for WPP. Horizontal lines correspond to results obtained with
the simple $\Delta k$ predictor. \emph{Middle row:} The rate
of correctly guessed new entries, $Q_n$ versus $\gamma$.
\emph{Bottom row:} Scatter plots of item degree increase in
$(t^*-T_P,t^*]$ versus the degree increase in $(t^*,t^*+T_F]$. Full
symbols correspond to items predicted by the WPP method.
In all cases, both history and future window lengths are $60$ days
for Netflix and Movielens and $10$ hours for Digg.}
\label{fig4}
\end{figure*}

In the PBP method, all users are considered equal: only the
number of users who have collected an item matters.
It is however possible that some users are better than the others in
detecting promising items and that their choice is only afterwards
followed by other users who are more popularity-driven and less
attentive to the emerging trends. However, Movielens and Netflix
datasets lack any additional user information which could allow us
to assess user weights.
We thus have to base our judgment only on the rating activity of
users which can be measured either as the number of recent ratings
$\Delta k_i(t^*,T_P)$ or as the total number of ratings $k_i(t^*)$.
Since the total activity performs slightly better in our
tests, we define the weighted popularity predictor (WPP) in the form
\begin{equation}
\label{weight}
s_{\alpha}(t^*,T_P)=
\sum_i\big(A_{i\alpha}(t^*)-A_{i\alpha}(t^*-T_P)\big)k_i(t^*)^{\gamma}.
\end{equation}
Here $\gamma$ is a tunable parameter which defines how much greater
(when $\gamma>0$) or lower (when $\gamma<0$) weight is given to active
users. When $\gamma=0$, the predicted score reduces to the popularity
increase in the past time window. As can be seen in Fig.~\ref{fig4},
prediction precision achieves a maximum around $\gamma=0$ in Movielens
and Netflix. In Digg, positive values of $\gamma$ lead to a considerable
increase in the precision value. Furthermore, both $\gamma>0$ and
$\gamma<0$ allow us to achieve significant rates of correctly predicted
new entries $Q_n$, which means that the method is able to detect promising
items. This feature is most pronounced when $\gamma>0$ in Digg. Notably,
approximately one third of these items are not found by the popularity-based
predictor because their popularity on the test date $t^*$ is too small.
As before, we make also scatter plots of the popularity increase (see
Fig.~\ref{fig4}) which further demonstrate the ability of the WPP to
detect the emerging items and avoid those that fade away. The
performance dependence on the future window length $T_F$ can be studied
too and shows that when $T_F$ increases, activity-favoring predictions
($\gamma>0$) suffer less than activity-disfavoring ones ($\gamma<0$).

\section{Trend Prediction Augmented by User Centrality in the Social Network}
We now focus on the Digg dataset for which, unlike the Movielens and
Netflix data, we have also the social network of connections among the
users. In the case of Digg, it is a directed leader-follower network thanks
to which followers receive stories ``dug'' by their leaders. This social
network contains 336,225 users and 2,251,171 links. The general idea is
that prominent users in the social network can be more effective
in propagating contents (because their reputation or their position in
the social network boost the propagation of a story) or have better
chances of digging promising contents (if their social status in the
social network reflects their ability to filter good stories). The
presence of influential users and their role in the propagation of
information is still a controversial subject and it is not clear to
which extent they can effectively influence the popularity of items
or products~\cite{WattsDuncanJandDodds2007}. This matter is also
debated in the context of viral
marketing~\cite{Leskovec2007,Subramani2003} where it is not clear if
large adoption of a product can be driven by a cascading word-of-mouth
process. Our data is not detailed enough to allow us to see if
prominent users (according to their number of followers or a more
sophisticated centrality measure) are in fact directly responsible
for propagation of stories. However, we can still assess if there is
some benefit to be gained in our prediction task from user status
in the social network.

We denote the adjacency matrix of the user social network as
$\mathsf{G}$. $G_{ij}=1$ if user $i$ follows user $j$
and $0$ otherwise. Since the network is directed, matrix $\mathsf{G}$
is not necessarily symmetric and we distinguish
between a node's in-degree (number of followers) $d^{IN}_i =
\sum_jG_{ji}$ and out-degree (number of leaders) $d^{OUT}_i =
\sum_j G_{ij}$. For computational reasons we do not take into account
the time dependence of the adjacency matrix $\mathsf{G}$ in the social
network. We denote the influence of user $i$ as $I_i$ (we specify it later)
and compute the influence-based predictor (IBP) similarly as in Eq.~(\ref{weight}),
that is by weighting the contribution of each user by this user's influence
\begin{equation}
\label{inf_weigth}
s_\alpha(t^*,T_P) = \sum_i \big(A_{i\alpha}(t^*)-
A_{i\alpha}(t^*-T_P)\big)I_i^\eta.
\end{equation}
Parameter $\eta$ makes it possible to tune the contribution of
user influence (when $\eta=0$, the method simplifies to the original
influence-free popularity increase).

We are free to choose from various influence measures (in social
sciences, the term centrality metric is often used instead~\cite{Bonacich1987}).
The simplest measure of influence is the user in-degree (the number of
followers), in which case we simply set
$I_i=d^{IN}_i$ and call the corresponding predictor IBP-IN. As more
refined measures of influence, we choose the PageRank~\cite{Page1999}
and the LeaderRank~\cite{Lu2011}, giving rise to two predictors:
IBP-PR and IBP-LR, respectively. Both PageRank and LeaderRank are
reputation metrics which derive the influence of a user from the
influence of his followers in a self-consistent way. These
two methods are shown to outperform the in-degree in identifying the
influential users for spreading~\cite{Lu2011}. In both algorithms,
users are first initialized with the same score. The PageRank score
of a user is then computed by iterating a process where a fraction
$\delta\in(0,1)$ of the score of a user is transferred in
equal shares to its leaders (we set $\delta=0.85$ as in the original
paper). The remaining $1-\delta$ fraction of the score is evenly
redistributed to all users in the network. The LeaderRank score is
computed in a similar way with the difference that $\delta$ is set
to $1$ and a ground node is introduced and connected with
all user nodes by bidirectional links.
This algorithm is parameter-free and it is based
on the assumption that users with few leaders owe a larger share of
their reputation to the entire community than users with many leaders.

\begin{figure}
\centering
\includegraphics[scale=0.46]{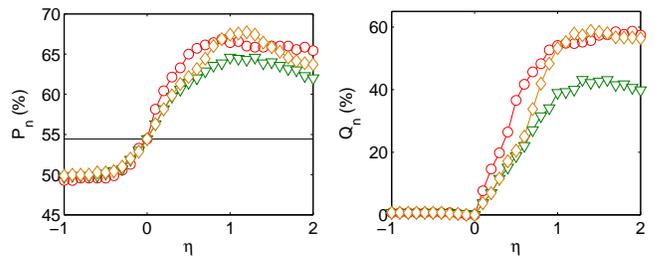}
\caption{(Color online) Prediction precision $P_n$ and
the rate of correctly guessed new entries $Q_n$ versus
$\eta$ in Digg for different user centrality measures: number of
followers (IBP-IN, red lines with circles), PageRank (IBP-PR, green
lines with triangles), and LeaderRank (IBP-LR, orange lines with
diamonds). Horizontal lines correspond to the results obtained with
the simple $\Delta k$ predictor.}
\label{fig6}
\end{figure}

Past work on predicting future popularity of Digg content
focuses on individual stories \cite{Szabo2010} and predictions of
the top-$L$ most popular items haven't been studied yet. We can thus only
present the results obtained with the simple $\Delta k$ predictor as
comparison. As one can see in Fig.~\ref{fig6}, all three measures of
influence yield significantly higher precision than the bare popularity
increase and the rate of correctly predicted new items is also
considerably higher than, for example, in Fig.~\ref{fig4}. Although
we cannot draw any causal implications from this observation, we can
say that measures of social influence importantly enhance the performance
of future popularity increase predictions. Performance obtained with
PageRank or LeaderRank weighting is better (with respect to both
$P_n$ and $Q_n$) than that obtained with bare in-degree,
confirming the added value of these two centrality measures. Note that
in the Digg dataset, a large part of these correctly guessed new entries
$Q_n$ cannot be predicted neither by their recent popularity
increase (by definition) nor by their total popularity. This means that
the IBP method is able to find inherently unexpected items whose
upcoming popularity is due to the social processes taking place in
the system. Similar results follow when top $n=50$ and $n=200$ places
of rankings are evaluated.

\section{Conclusions}
We investigated the ability of different methods to predict which items
are going to have the biggest popularity increase in the near
future. When items in the studied system have short typical lifetime
(which, for example, is the case for the Digg data studied herein),
predictions by total popularity result in poor performance while
predictions by recent popularity perform well. In Netflix and Movielens
data we find that recent popularity is a good
predictor for short future time windows but its performance decreases
fast with the future window length. Predictions by total popularity,
instead, are more stable in this sense and perform reasonably well
also in the long run. By combining these two predictors, one
can achieve a slightly higher precision and a large increase in the
number of correctly guessed items that are new at the top of the ranking.
We found in all studied datasets that weighted popularity increase
which takes user activity into account, while not so useful for
improving the prediction precision, can detect items whose popularity
was not particularly high in the recent past. Finally, in the case of
the Digg data we found that knowledge of the underlying user social
network can significantly enhance the prediction results. To achieve
this improvement, we weighted users with various measures of social
status (in-degree, PageRank, and LeaderRank) and found that both
precision and the ability to predict items that were recently not so
popular improve. In summary, the hybrid method combining the
total item popularity with recent popularity allows for some improvements
in the case of Netflix and Movielens. In Digg, the benefit gained from
the knowledge of the social network among the users is substantial and
the weighted predictor based on social influence achieves improvements
in accuracy and, even more, in the ability to correctly predict new
items at top places of the ranking.

Our study is an exploratory one and there is much work that remains to
be done in the future. First of all, to test the methods on
more datasets would be useful to show possible limits of their
applicability. One should also invest the computational effort to
evaluate the methods on large-scale data---both for the sake of
confirming previously found patterns and for evaluating which
computation steps can be simplified. For practical applications of
the ideas proposed in this work, it would be very important to devise
scalable algorithms able to cope with the massive data routinely
produced by the current online systems. For large-scale
data, one can also devise methods that benefit from the
often-available additional information such as user and item meta data.
Robust statistical techniques could reveal that, for example, users
with certain background (say, females under 25 years) are particularly
significant for predicting popularity of a specific kind of contents.
How much this could improve the predictions is of course an open
question.

Besides devising further techniques for trend prediction, we find it
important to search for additional metrics to assess the prediction
performance. For example, incorporating the order of the
ranking will increase the information of the prediction. Moreover,
it would be interesting to focus on items
which are in the early stage of their evolution---one can say that the
ability to predict success of those would be of foremost usefulness to
the users. Our metric $Q_n$ makes a step in this direction by counting
the items which are new in the top-$100$ ranking of items by their recent
popularity. However, it does not account for the fact that some of
those ``new'' items can already have substantial total popularity and
they only return to the group of recently popular items after a
momentous lapse. The natural way to aim for those well-performing new
items is to define the logarithmic derivative of the popularity,
$\Delta k/k$, as the true score and see how to predict this ranking.
However, we found it difficult to work with $\Delta k/k$ because of
the excess weight that it puts on low-degree items (highest values
are achieved by items with very low $k$) and the resulting sensitivity
to the discreteness of time in the data. For example, an item
submitted short before midnight accumulates only a few links on the
first day and then excels---to some extent without reason---in
$\Delta k/k$ the day after. Devising more reliable and justifiable
metrics focusing on genuinely new items thus remains a future
challenge.

\section*{Acknowledgments}
This work was partially supported by the Future and Emerging
Technologies program of the European Commission FP7-COSI-ICT (project
QLectives, grant no. 231200) and by the Swiss National Science
Foundation (grant no. 200020-132253).

\bibliographystyle{ws-acs}
\bibliography{tp}

\end{document}